\documentclass[conference]{IEEEtran}
\IEEEoverridecommandlockouts
\usepackage{cite}
\usepackage{amsmath,amssymb,amsfonts}
\usepackage{algorithmic}
\usepackage{graphicx}
\usepackage{textcomp}
\usepackage{xcolor}
\usepackage{enumitem}
\usepackage{multirow}
\usepackage{booktabs}
\usepackage{hyperref}
\def\BibTeX{{\rm B\kern-.05em{\sc i\kern-.025em b}\kern-.08em
    T\kern-.1667em\lower.7ex\hbox{E}\kern-.125emX}}
\begin{document}

\title{
% Text-Guided Fine-Grained \\ Audio Generation with Diffusion Transformer
% Text-Only Driven Fine-Grained \\ Audio Generation with Diffusion Transformer
% Towards Fine-grained Audio Generation with Natural Language Descriptions Based on Flow-based Diffusion Transformers
AudioComposer: Towards Fine-grained Audio Generation with Natural Language Descriptions
% Audio Generation from Control-Rich Text Inputs\\
% {\footnotesize \textsuperscript{*}Note: Sub-titles are not captured for https://ieeexplore.ieee.org  and
}

\author{
% \begin{tabular}{c}
% Yuanyuan Wang$^{1}$, Hangting Chen$^{2,\dagger}$, Dongchao Yang$^1$, Xixin Wu$^{1,\dagger}$, Zhiyong Wu$^{1,3}$, Helen Meng$^1$
% \end{tabular}
\IEEEauthorblockN{
Yuanyuan Wang$^{1, \ast}$, Hangting Chen$^{2,\dagger}$, Dongchao Yang$^1$, Zhiyong Wu$^{1,3}$, Xixin Wu$^{1,\dagger}$
}
% \thanks{$^{\ast}$Work performed during an internship at Tencent AI Lab.} 
% \thanks{$^\dagger$Corresponding author.}

\IEEEauthorblockA{$^1$ The Chinese University of Hong Kong, Hong Kong SAR, China} 

\IEEEauthorblockA{$^2$ Tencent AI Lab, Audio and Speech Signal Processing Oteam, China}

\IEEEauthorblockA{$^3$ Shenzhen International Graduate School, Tsinghua University, Shenzhen, China} 

}

% \and
% \and
% \and
% \and
% \and

\maketitle

\begin{abstract}
% Text-to-audio (TTA) generation aims to generate corresponding audios based on text descriptions. 
%挑战1: data scarcity：
Current Text-to-audio (TTA) models mainly use coarse text descriptions as inputs to generate audio, which hinders models from generating audio with fine-grained control of content and style.
%挑战2: complex：
Some studies try to improve the granularity by incorporating additional frame-level conditions or control networks. 
% in addition to the utterance-level text descriptions, and leveraging complex control networks to extract the control information from the frame-level conditions. 
However, this usually leads to complex system design and difficulties due to the requirement for reference frame-level conditions.
To address these challenges, we propose AudioComposer, a novel TTA generation framework that relies solely on natural language descriptions (NLDs) to provide both content specification and style control information. 
% which significantly improves generation quality and controllability.
%simplicity and efficiency：
To further enhance audio generative modeling, we employ flow-based diffusion transformers with the cross-attention mechanism to incorporate text descriptions effectively into audio generation processes, which can not only simultaneously consider the content and style information in the text inputs, but also accelerate generation compared to other architectures. 
% data scarcity：
Furthermore, we propose a novel and comprehensive automatic data simulation pipeline to construct data with fine-grained text descriptions, which significantly alleviates the problem of data scarcity in the area.
Experiments demonstrate the effectiveness of our framework using solely NLDs as inputs for content specification and style control. The generation quality and controllability surpass state-of-the-art TTA models, even with a smaller model size.
\footnote{Demo and code are available in
\href{https://lavendery.github.io/AudioComposer/}{https://lavendery.github.io/AudioComposer/}.
$^{\ast}$Work performed during an internship at Tencent AI Lab.
$^\dagger$Corresponding author.
This work was supported by National Natural Science Foundation of China (62306260,62076144).\label{demo}}
\end{abstract}


\section{Introduction}

Text-to-audio (TTA) generation focuses on generating authentic and accurate audios corresponding to the information specified in text inputs~\cite{yang2023diffsound}. 
TTA plays a crucial role in producing various sound effects and applies to diverse fields including movie sound effect creation, virtual reality, game design, audio editing, and interactive systems~\cite{huang2023makeanaudio2, guo2024audio}.
In recent years, there have been remarkable advancements in deep generative models~\cite{kingma2018glow, goodfellow2020generative, ho2020denoising}, which have substantially contributed to the development of audio generation.
Some recent works have made considerable progress by employing diffusion models~\cite{yang2023diffsound, huang2023makeanaudio, liu2023audioldm, ghosal2023texttoaudio, ghosal2023tango, audioldm2-2024taslp, majumder2024tango} or autoregressive models~\cite{kreuk2023audiogen, yang2023uniaudio, yang2024uniaudio15}.
% for generating audio conditioned on given textual prompts. 
These existing approaches primarily concentrate on audio generation based on coarse text content descriptions, which in turn restricts the style controllability of generating fine-grained audio. 
For instance, they are unable to specify and generate accurate temporal locations of sound events, which is a crucial limitation for applications such as video dubbing. 

To achieve such detailed control, the availability of fine-grained text-audio pair data is a vital prerequisite. Nevertheless, such paired text-audio data is generally difficult to obtain, especially with fine-grained instructions.
Recently, Make-An-Audio2~\cite{huang2023makeanaudio2} employed large language models (LLMs) to augment structured captions into natural language captions, which alleviates the issue of insufficient temporal paired data. 
However, this approach is constrained to the simple instructions with temporal orders (e.g., ``\textit{sound A
and then sound B}") and fails to specify more detailed information, like precise timestamps and durations (e.g., ``\textit{sound A starts from 2.5 seconds and lasts for 3 seconds}"). 
Another research line for improving fine-grained controllability proposes to incorporate extra control conditions into the TTA systems.
PicoAudio~\cite{xie2024picoaudio} utilized frame-level timestamp information as complementary conditions to text inputs. 
Guo~\textit{et al.}~\cite{guo2024audio} designed a specialized encoder to extract control information from frame-level conditions and a Fusion-Net for integrating the fine-grained control information in the generation process. 
However, these approaches increase model complexity and also bring difficulties during inference as extra frame-level conditions from reference audios are required.
Therefore, current controllable TTA research still faces three main challenges:
(1) \textbf{scarcity} of fine-grained audio-text data; 
(2) \textbf{complexity} in incorporating control information; 
(3) lack of \textbf{precision} in control capabilities.

\begin{figure*}[!t]
\centering
\includegraphics[scale=0.8]
{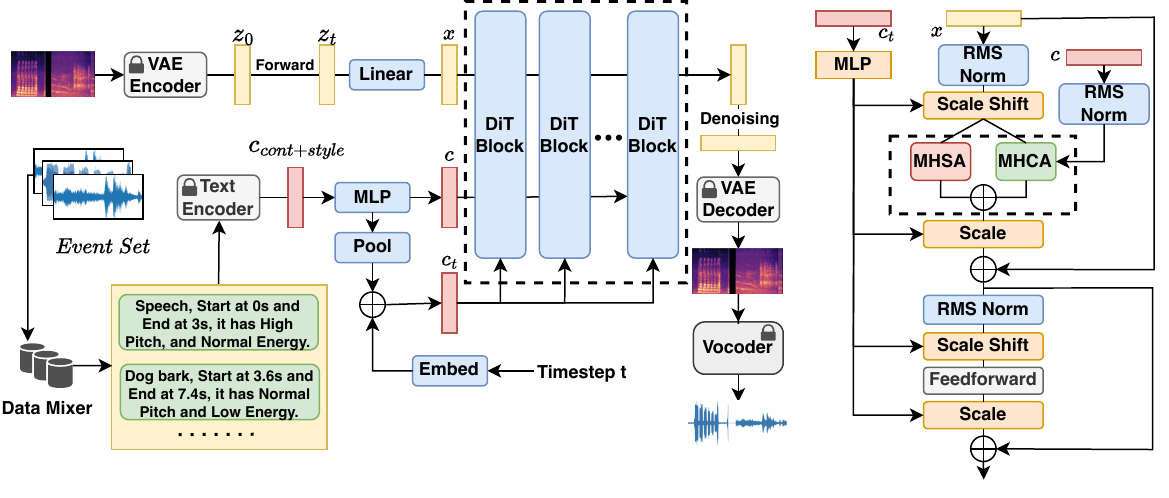}
\vspace{-17pt}
\caption{The left picture shows the overview of AudioComposer, while the right picture illustrates the details of DiT Block. MHSA: Multi-Head Cross-Attention. MHCA: Multi-Head Self-Attention. $\oplus$ means add.}
\vspace{-17pt}
\label{fig:overall}
\end{figure*}
In this paper, we propose AudioComposer, a fine-grained audio generation framework based on flow-based diffusion transformers with only text as inputs for both content specification and style control. 
First, to mitigate the issue of data scarcity, we introduce an innovative online data simulation pipeline that enables fine-grained style annotations, including information of timestamps, pitch, and energy, with natural language descriptions (NLDs). 
Second, leveraging the automatic data simulation pipeline, AudioComposer exclusively depends on fine-grained natural language guidance for controllable TTA generation. This approach eliminates the need for additional frame-level conditions or complex control networks, achieving high simplicity and efficiency. 
Finally, inspired by the success of flow matching and diffusion transformers~\cite{lipman2022flow, peebles2023scalable}, considering the need to preserve fine-grained text representation~\cite{esser2024scaling, gao2024lumi}, we explore flow-based diffusion transformers by integrating text conditions through cross attention mechanisms, which can capture inherent connections between fine-grained text representation and latent audio tokens.
This architecture not only accelerates the generation process but also enhances audio generative performance. 
In summary, the main contributions of this paper are as follows:

\begin{itemize}[itemsep=0pt,topsep=0pt,parsep=0pt,leftmargin=10pt]
\item 
% 1.
We present a comprehensive automatic data simulation pipeline to generate fine-grained NLDs, which effectively tackles the issue of data scarcity in controllable TTA systems. 
\item 
% 2.
Our method utilizes NLDs to enable precise control in TTA generation, eliminating the need for additional conditions or complex control networks.
\item 
% 3.
We employ flow-based diffusion transformers with the cross-attention mechanism, 
% which accelerates the generation process and achieves more precise and fine-grained control performance.
which improves generation speed, quality, and controllability.
\end{itemize}

% First, to address the fine-grained temporal disorder, we leverage a temporal condition encoder to take the timestamp as input.
% In this way, we alleviate the pressure on the text encoder to identify events along with their corresponding temporal information, by allowing the T2A system to model time information of events more efficiently.
% Secondly, we provide flexible timestamps. 
% During training, we generate mixed audio and their timestamps online without restricted timestamps input. 
% During inference, we use LLM to supply formatted timestamps without manual input, making the process highly convenient. 
% Thirdly, to tackle the challenge of insufficient temporally aligned audio-text paired datasets, we conduct data augmentation online, which utilizes publicly available datasets to generate new mixed audio online. 
% This approach allows for dynamic provision of timestamps and is not limited to a specific dataset, resulting in a more diverse range of audio data.

\section{Methodologies}
% Our TTA generation framework, as shown in Fig.\ref{fig:overall}, consists of a text encoder to extract information from the data mixer as conditions, a diffusion model to generate latent representations based on the conditions, and a variational auto-encoder (VAE) to predict the mel-spectrograms that are further reconstructed to waveform by a vocoder.
This section presents our fine-grained TTA generation system AudioComposer. 
The overall architecture is demonstrated in Fig.~\ref{fig:overall}, consiting of a variational auto-encoder
(VAE)~\cite{kingma2013auto}, a flow-based diffusion transformer (DiT) model~\cite{peebles2023scalable}, a text encoder~\cite{raffel2020exploring}, and a vocoder~\cite{lee2022bigvgan}. 
Following previous works \cite{huang2023makeanaudio2,huang2023makeanaudio}, we use pre-trained and frozen Mel-VAE and BigVGAN-based vocoder~\cite{lee2022bigvgan} in our framework.
%For the text encoder, we adopt pre-trained T5-large model. 
%Note that the Mel-VAE, vocoder and text encoder are frozen during training. 
% We give the details of the proposed AudioComposer and automatic data simulation pipeline in Section \ref{flow-based diffusion} and \ref{data mix}.
% we utilize a pre-trained BigVGAN~\cite{lee2022bigvgan} as our vocoder.
% As shown in Fig.\ref{fig:overall}, we first transform a sequence of sound event sets into fine-grained text descriptions and then mixed audio using our data mixer. 
% For these text descriptions, they are fed into a text encoder to extract the corresponding text representation, denoted as $c_{desc+style}$. 
% For the waveforms, they are converted into mel-spectrograms, which are then input into the VAE Encoder. The VAE Encoder outputs a compressed latent variable $z_0$. 
% After that, $z_0$ and $c_{desc+style}$ are processed through our flow-based diffusion transformers. 
% The output of our model is then converted into a mel-spectrogram by the VAE decoder, which is ultimately transformed into a waveform by the vocoder.
% A.  The details of automatic data simulation pipeline
% B. The fine-grained TTA framework
% B.1: overview
% B.2: flow-based diffusion 
% B.3 Cross attention
%\subsection{The fine-grained Audio Generation Framework}
% \subsubsection{Overview}

\subsection{Data Mixer for AudioComposer} \label{data mix}
\label{sec:datamixer}
Our data mixer can generate a dataset, referred to as AudioTPE, with fine-grained annotations which include event labels, timestamps, pitch, and energy information.
The simulation process to obtain AudioTPE consists of the following steps.
% It can process three types of input data:
% \textbf{Clean Event}. 
First, for datasets comprising a series of clean sound events, where typically each audio contains only one sound event, we initially calculate the average pitch and energy for each audio file, which are partitioned into high, normal, and low categories based on the 25\%, 50\%, and 75\% quantiles. 
In this way, each audio possesses its distinct pitch and energy categories. We then randomly select these annotated audios to simulate mixed audios with durations less than 10 seconds. During this process, we record start time, end time, pitch category, and energy category of each event in simulated audios, which yields fine-grained annotated audio data.
Finally, we generate NLDs based on these annotations using a template, e.g., ``\textit{Dog bark, Start at 3.6s and End at 7.4s, it has Normal Pitch and Low Energy.}''

\subsection{AudioComposer} \label{flow-based diffusion}
Flow matching~\cite{lipman2022flow, gao2024lumi} has demonstrated powerful generation performance and efficiency in image processing fields. In this paper, we explore the ability of flow-based diffusion models in AudioComposer. Inspired by DiT \cite{peebles2023scalable,esser2024scaling,simplespeech,simplespeech2} and Sora, we also propose to combine flow-based diffusion formulation and transformer-based structure. In the following, we first review the flow matching and then
show how to achieve fine-grained control using only NLDs. 
% show the details of the diffusion transformer structure and the application to TTA in style control. 
\subsubsection{Conditional Flow Matching}
% In this paper, we utilize a flow-based diffusion transformer. 
% Therefore, we first revisit the methodology of flow matching for generative models~\cite{lipman2022flow, gao2024lumi}, which performs a linear interpolation between noise and data along a straight line.
%Following the definition of 
Conditional flow matching (CFM) \cite{lipman2022flow} aims to learn a mapping between samples $\varepsilon \sim \mathcal{N}(0, I)$ from noise distribution to samples $x \sim p(x)$ from data distribution through the following interpolation-based forward process between the time interval $[0,1]$:
\vspace{-2pt}
\begin{equation}\label{eq1}
\begin{aligned}
x_t = \alpha_tx + \beta_t\epsilon, 
\end{aligned}
\vspace{-2pt}
\end{equation}
where $\alpha_0 = 0, \beta_0 = 1, \alpha_1 = 1$, and $\beta_1 = 0$. With different choices of $\alpha_t$ and $\beta_t$, different interpolation schedules are obtained, e.g., the linear interpolation $x_t = tx + (1-t)\epsilon$ with $\alpha_t=t, \beta_t=1-t$. The goal is to employ a trainable neural network $v_t(x_t; \theta)$ to approximate the time-dependent velocity field $u_t(x_t) = \alpha_t^{\prime}x + \beta_t^{\prime}\epsilon$, where $\alpha_t^{\prime}$ and $\beta_t^{\prime}$ denote the derivatives of $\alpha_t$ and $\beta_t$ with respect to time t. The training loss can be defined as~\cite{gao2024lumi}:
\begin{equation}\label{eq6}
\begin{aligned}
\mathcal{L}_{\tt flow} = \mathbb{E}_{t\sim\mathcal{U}[0,1],\epsilon\sim\mathcal{N}(0, I),p(x)}||v_t(x_t; \theta) - u_t(x_t)||^2,
\end{aligned}
\end{equation}
where $\mathcal{U}[0,1]$ is a uniform distribution, sharing similarity with the noise prediction or score prediction losses in diffusion.

With the trained network, we can transform noise samples into data samples by solving the flow ordinary differential equation (ODE) from $t=0$ to $t=1$:
\vspace{-2pt}
\begin{equation}\label{eq5}
\begin{aligned}
\mathrm{d}x_t = v_t(x_t; \theta)\mathrm{d}t.
\end{aligned}
\vspace{-2pt}
\end{equation}

% \subsubsection{Diffusion Transformer}
% \label{diff_trans}
% DiT~\cite{peebles2023scalable} explores replacing the U-Net backbone with a transformer that operates on latent patches, attains state-of-the-art FID scores on class-conditional ImageNet benchmarks, and showcases superior scaling capabilities. 
% In this work, our architecture is based on the DiT~\cite{peebles2023scalable}. 

\subsubsection{Fine-grained Control with Natural Language Descriptions}

% As illustrated in Section~\ref{diff_trans}, besides simply incorporating conditioning via adaptive layer norm for generation, our DiT blocks can flexibly support text instruction with zero-initialized attention~\cite{zhang2023llama, gao2023llama, zhang2023adding} by incorporating fine-grained text representation $c$. 
% Additionally, we replace all LayerNorm~\cite{ba2016layernormalization} with RMSNorm~\cite{zhang2019root} to enhance training stability.
% The inputs of DiT Block contain noisy latent tokens $x$, fine-grained text representation $c$, and modulation representation $c_t$. 
To ensure capturing content specification and style control information, we utilize the pre-trained T5~\cite{raffel2020exploring} as the text encoder based its outstanding natural language understanding capability. The content and style information is extracted by T5 from NLDs into the representation $c_{cont+style}$. Based on the pre-trained VAE, $z_0$ is generated from mel-spectrograms. The diffusion process is conducted to gradually add noise to $z_0$ to obtain the noisy tokens $z_t$, which is further encoded to a latent representation $x$ using linear layers.
As depicted in the left picture of Fig.~\ref{fig:overall}, we employ latent representations $x$ and text representations of $c$ and $c_t$ as inputs to the DiT blocks.
Following the standard DiT that utilizes a modulation mechanism~\cite{esser2024scaling} to condition the network on both the timesteps of the diffusion process and the class labels, we combine embeddings of the timestep $t$ and pooled representations of $c_{cont+style}$ into $c_t$ as inputs for the DiT blocks. 
Considering that the pooled text representations retain merely coarse-grained information about text inputs~\cite{podell2023sdxl}, while our AudioComposer needs fine-grained text representations to achieve accurate control~\cite{gao2024lumi, esser2024scaling}, we also incorporate the sequence representation $c$ transfromed from $c_{cont+style}$ by multi-layer perceptron (MLP) without pooling into each DiT block. 

The structure of the DiT blocks is shown in the right picture of Fig.~\ref{fig:overall}.
% In the cross-attention process, 
The queries $x_q$ of latent audio tokens are used to aggregate information from keys and values of text representations $c$ using a cross-attention mechanism, which helps capture inherent connections between text and latent audio tokens and contributes to fine-grained information extraction. 
% Subsequently, we introduce a zero-initialized gating mechanism to inject conditional information into the token sequences gradually.
Given audio queries $x_q$, keys $x_k$, and values $x_v$ with text keys $c_k$ and values $c_v$, the final output of self-attention and cross-attention is computed as:
\vspace{-3pt}
\begin{equation}\label{eq7}
\begin{aligned}
A = softmax\left(\frac{\widetilde{x}_q\widetilde{x}_k^T}{\sqrt{d}}\right)x_v + tanh(\alpha)softmax\left(\frac{\widetilde{x}_q c_k^T}{\sqrt{d}}\right)c_v, \nonumber
\end{aligned}
\vspace{-3pt}
\end{equation}
where $\widetilde{x}_q$ and $\widetilde{x}_k$ means applying Rotary Position Embedding (RoPE)~\cite{su2024roformer} to audio queries and keys, $d$ represents the dimension of queries and keys, $\alpha$ denotes the zero-initialized learnable parameter in the gated cross-attention~\cite{zhang2023llama, gao2023llama, zhang2023adding}. 
Using the outputs of the last DiT block, we can recover the latent tokens by solving ODE during inference. Lastly, we can get waveforms with the help of VAE decoder and Vocoder.
% The VAE decoder is then utilized to predict the mel-spectrogram, which is further reconstructed to waveform by the vocoder.

% \textbf{AudioCondition}~\cite{guo2024audio}. The AudioCondition dataset is derived from the AudioSet strong label data, which already incorporates timestamp annotations. However, the sound events within this dataset are noisy and may encompass multiple events within a specific duration. Therefore, the pitch and energy calculations cannot accurately represent individual events, and we solely control the precise timestamp in this case.

% \textbf{AudioCaps}~\cite{audiocaps}. Employing text exclusively for fine-grained control offers the advantage of utilizing other coarse-grained data, such as AudioCaps, to enhance generation performance. As a result, our data mixer and model can leverage text descriptions to accomplish the audio generation task without fine-grained control. We will examine its impact on model performance in our experiments.

\section{experimental setup}
% \vspace{-10pt}
\subsection{Dataset}
\label{sec:dataset}
In the experiments, we use three types of datasets:
(1) \textbf{AudioTPE data}. We combine several datasets, including FSD50K~\cite{fonseca2021fsd50k}, ESC50~\cite{piczak2015dataset}, UrbanSound8K~\cite{salamon2014dataset}, ODEON{\_}Sound{\_}Effects\footnote{https://www.paramountmotion.com/odeon-sound-effects}, to simulate mixed data online. Each audio in the raw datasets typically encompasses a single sound event.
With a total of 49k audios amounting to approximately 140 hours, we perform online simulation and fine-grained annotation of the data, as introduced in Section~\ref{sec:datamixer} to obtain the final AudioTPE dataset that contains fine-grained annotations on information of events, time, pitch, and energy.
(2) \textbf{AudioCondition}~\cite{guo2024audio}. This dataset is derived from AudioSet Strong~\cite{45857,9414579}, which contains about 81k audios, in total about 230 hours, with temporally strong labels. 
(3) \textbf{AudioCaps}~\cite{audiocaps}. 
This is a large-scale dataset of about 46K audio clips to human-written text pairs, totaling around 120 hours. 
It contains only content information without style control.
% Employing text solely for fine-grained control offers the benefit of also being able to utilize other coarse-grained data to enhance generation performance. 
% Our AudioComposer can also be trained on AudioCaps without fine-grained descriptions. %as verified in Section~\ref{sec:ablation}.

% (3) We directly utilize the AudioCaps~\cite{audiocaps} dataset, which contains 45k audio clips paired with human-written captions, to train our text-to-audio generation model without fine-grained control for improving generation performance. 

\subsection{System Configuration}
We train the VAE to compress mel-spectrograms into 20-dimension latent representations. 
The AudioComposer is trained on 8 NVIDIA V100 GPUs, using a batch size of 16 per GPU.
We employ the AdamW optimizer~\cite{kingma2014adam} with a learning rate of 1e-4. 
All model is trained with 70k steps, and the transformer head is 32. 
The AudioComposer has two versions with different parameter sizes, i.e., AudioComposer-S with 6 transformer blocks and a hidden size of 768, and AudioComposer-L with 9 transformer blocks and a hidden size of 1024.

\begin{table}[tb]
\centering
\setlength{\tabcolsep}{1.5mm}
\caption{Results on AudioCondition Test Set. -S: Small, -L: Large.}
\vspace{-6pt}
\begin{tabular}{cccccc}
\toprule[1pt]
\multirow{2}{*}{\textbf{Model}} 
&
\multirow{2}{*}{\textbf{\#Params}}
& \multicolumn{2}{c}{\textbf{Objective$(\%)\uparrow$}}
& \multicolumn{2}{c}{\textbf{Subjective$\uparrow$}}
\\
&
      & \multicolumn{1}{c}{$F1_{event}$} & \multicolumn{1}{c}{$F1_{seg}$} & \multicolumn{1}{c}{$MOS_t$} 
    & \multicolumn{1}{c}{$MOS_q$}
      \\
\hline
Ground Truth
& \text{-}
& 43.36
& 63.46
& 4.01
& 4.24
% \\
% PicoAudio~\cite{xie2024picoaudio}
% & \text{-}
% & 8.33 & 25.48 & \text{-} & \text{-}
\\
AudioLDM-L-Full
 & 739M
& 3.21           & 23.94        & \text{-} & \text{-}
\\
AudioLDM2
 &  346 M
& 4.71           & 39.72        & \text{-} & \text{-}
\\
AudioLDM2-Large
 & 712M
& 8.4          & 46.19        & \text{-} & \text{-}
\\
Tango
 & 866M
& 1.6         & 26.51        & \text{-} & \text{-}
\\
Tango2
 & 866M
& 4.04           & 39.41        & 1.69 & 2.8
\\
MC-Diffusion~\cite{guo2024audio}
 & 1076M
& 29.07           & \text{-}        & \text{-} & \text{-}
\\
\hline
Tango+LControl  
&
866M
& 
21.46                                    &  55.15   & 2.47
& 3.01
\\
% Tango+FControl 
% &
% 866M
% & 
% xxx                                   &  xxx    
% & 2.41
% & 3.11
% \\
% Tango+LControl-Full
% &
% 866M
% & 
% 13.59                                    &  56.35    & xxx
%  & xxx
% \\
\textbf{AudioComposer-S}
&
272M
& 
\underline{43.51}                                    &  \underline{60.83}                                 & \textbf{4.6}
& \underline{3.81}
\\
\textbf{AudioComposer-L}
&
742.79M
& 
\textbf{44.4}                                    &  \textbf{63.3}                                
& \underline{4.51}
& \textbf{4.02}
\\
\bottomrule[1pt]
% \multicolumn{6}{l}{$^{*}$S: Small, L: Large}
% \\
\end{tabular}
\label{res_on_audiocondition}
\vspace{-18pt}
\end{table}

\begin{table*}[tb]
\centering
\setlength{\tabcolsep}{1.7mm}
\caption{Results on AudioTPE Test Set. -S denotes the small model size, and -L denotes the large model size.}
\vspace{-7pt}
\begin{tabular}{ccccccccccc}
\toprule[1pt]
\multirow{2}{*}{\textbf{Model}} 
&
\multirow{2}{*}{\textbf{\#Params}}
& \textbf{Timestamp}
& \multicolumn{2}{c}{\textbf{Pitch}}
& \multicolumn{2}{c}{\textbf{Enenrgy}}
& \multicolumn{4}{c}{\textbf{Subjective$\uparrow$}}
\\
&
      & \multicolumn{1}{c}{$F1_{seg}$$(\%)\uparrow$} & \multicolumn{1}{c}{$ACC(\%)\uparrow$} & \multicolumn{1}{c}{$MAE\downarrow$} 
    & \multicolumn{1}{c}{$ACC(\%)\uparrow$} & \multicolumn{1}{c}{$MAE\downarrow$} & \multicolumn{1}{c}{$MOS_t$} 
    & \multicolumn{1}{c}{$MOS_p$}  
    & \multicolumn{1}{c}{$MOS_e$} 
    & \multicolumn{1}{c}{$MOS_q$}
      \\
\hline
Ground Truth
& \text{-}
& 60.63
& 78.75
& \text{-}
& 90.16
& \text{-}
& 4.57
& 4.41
& 4.4
& 4.48
\\
AudioLDM-L-Full
 & 739M
& 23.11          & 29.98        & 108.97 & 37.81
& 34.4
& \text{-}
& \text{-}
& \text{-}
& \text{-}
\\
AudioLDM2
 &  346 M
& 41.54           & 33.55       & 129.05 & 43.62
& 36.61
& \text{-}
& \text{-}
& \text{-}
& \text{-}
\\
AudioLDM2-Large
 & 712M
& 39.72          & 33.56       & 118.87 & 42.05
& 34.56
& \text{-}
& \text{-}
& \text{-}
& \text{-}
\\
Tango
 & 866M
& 33.79        & 35.57        & 116.95 & 44.07
& 30.11
& \text{-}
& \text{-}
& \text{-}
& \text{-}
\\
Tango2
 & 866M
& 44.35          & 34.45       & 118.6 & 48.99
& 30.7
& 2.11
& 2.59
& 2.61
& 2.77
\\
\hline
Tango+LControl
&
866M
& 
47.91                                   &  39.37    & 113.9
 & 52.13
 & \textbf{27.4}
 & 4.14
& 3.41
& \underline{3.93}
& 3.68
\\
\textbf{AudioComposer-S}
&
272M
& 
\underline{50.97}                                    &  \textbf{60.63}                                
& \underline{91.25}
& \underline{63.53}
& 37.33
& \underline{4.5}
& \textbf{3.71}
& 3.63
& \underline{3.82}
\\
\textbf{AudioComposer-L}
&
742.79M
& 
\textbf{51.36}                                    
& \underline{56.6}                               
& \textbf{87.72}
& \textbf{65.77}
& 37.03
& \textbf{4.58}
& \underline{3.64}
& \textbf{4.20}
& \textbf{4.11}
\\
\bottomrule[1pt]
\end{tabular}
\label{res_on_cleanevent}
\vspace{-18pt}
\end{table*}

\begin{table}[tb]
\centering
\setlength{\tabcolsep}{1.5mm}
\caption{Ablation Studies on AudioCondition Test Set.}
\vspace{-6pt}
\begin{tabular}{ccc}
\toprule[1pt]
\textbf{Model}
& $F1_{event}(\%)\uparrow$
& $F1_{seg}(\%)\uparrow$
\\
\hline
Ground Truth
& 43.36
& 63.46
\\
\textbf{AudioComposer-S}
& \textbf{43.51}          & \textbf{60.83} 
\\ 
\hline
w 200 inference steps
& 46.43 & 66.19
\\
% \\
% \hline
w/o flow matching
& 35.19        & 46.92 
\\
w/o AudioCaps
& 34.55          & 47.84      
\\
\bottomrule[1pt]
\end{tabular}
\label{res_on_ablation}
\vspace{-17pt}
\end{table}

\subsection{Evaluation Metrics}
% In our experiments, both objective and subjective evaluation metrics to perform comprehensive assessments.

\noindent{\textbf{Objective metrics.}}
For timestamp control, we employ a sound event detection (SED) model to pinpoint the locations of generated sound events. The open-source SED system PB-SED\footnote{https://github.com/fgnt/pb\_sed} ~\cite{ebbers2022pre} is used. 
We utilize event-based macro F1-score ($F1_{event}$), and segment-based macro F1-score ($F1_{seg}$) to evaluate event accuracy~\cite{mesaros2016metrics}. 
%ranked first in DCASE 2022 Task 4,
% For pitch and energy control, given that both are categorized into three types: high, normal, and low, we initially compute the mean pitch and energy for the event segment within each audio clip.
% Subsequently, these audios can be classified based on predefined thresholds, and results are compared with our pre-annotated categories. 
% Finally, this enables us to compute the accuracy of pitch and energy across all audio clips. 
For pitch and energy control, we evaluate the accuracy ($ACC$) of pitch and energy categories across all audio clips. 
We also evaluate the mean absolute error (MAE) between frame-wise pitch and energy extracted from generated audios and those from ground-truth audios~\cite{ren2020fastspeech}.

\noindent{\textbf{Subjective metrics.}}
We conduct mean opinion score (MOS) assessment from multiple perspectives:
(1) temporal controllability ($MOS_t$) for evaluating accuracy of timestamp control;
(2) pitch controllability ($MOS_p$) for assessing accuracy of pitch control;
(3) energy controllability ($MOS_e$) for measuring accuracy of energy control, and
(4) audio quality ($MOS_q$) for evaluating naturalness of generated audios (without taking into account the accuracy of time, pitch, or energy).
For each task, 10 test groups from each model are rated by 10 evaluators, and the mean score is calculated.

\section{Results and Analyses}
\subsection{Results on AudioCondition}
\label{sec:res_on_audiocondition}

In this experiment, we train AudioComposer on all datasets introduced in Section~\ref{sec:dataset} and evaluate it on AudioCondition. 
We compare AudioComposer with mainstream generative models, including AudioLDM~\cite{liu2023audioldm}, AudioLDM2~\cite{audioldm2-2024taslp}, Tango~\cite{ghosal2023tango}, Tango2~\cite{majumder2024tango}, and MC\text{-}Diffusion~\cite{guo2024audio}, to assess the performance of temporal controllability, with the results provided in Table~\ref{res_on_audiocondition}. 
% However, the sound events within this dataset are noisy and may encompass multiple events within a specific duration. Therefore, the pitch and energy calculations cannot accurately represent individual events, and we solely control the precise timestamp in this case. 
As Tango and AudioLDM generate audio without the need for temporal conditions, we train Tango \footnote{https://github.com/declare-lab/tango/tree/master} from scratch with language control as a more direct comparison baseline, namely Tango+LControl. 
% \textcolor{red}{This provides a more direct comparison baseline in Table~\ref{res_on_audiocondition} and ~\ref{res_on_cleanevent}.}
Considering the lack of style control in open-source models and the cost of human resources, we only select Tango2 for the MOS evaluation. 

Table~\ref{res_on_audiocondition} shows that our AudioComposer outperforms these baseline models across all metrics. 
Interestingly, AudioComposer even surpasses the ground truth in $F1_{event}$ and $MOS_t$. This can be attributed to the fact that the ground truth is derived from AudioSet, which contains some extraneous noise apart from specific events.
% These results collectively illustrate the outstanding performance achieved by AudioComposer, even with a smaller number of parameters in AudioComposer-S.  
These results demonstrate the outstanding performance of AudioComposer, even with fewer parameters in AudioComposer-S.

% Moreover, we compare the language-guided and frame-level control methods using $Tango+LControl$ (Language-guided Control) and $Tango+FControl$ (Frame-level Control method).
% The frame-level Control method is trained with frame-level timestamps into the diffusion process, similar to PicoAudio~\cite{xie2024picoaudio}. 
% The results of $Tango+LControl$ and $Tango+FControl$ are quite similar, which can validate the effectiveness of Language-guided Control. 

\subsection{Results on AudioTPE}
% \vspace{-2pt}
Similar to Section~\ref{sec:res_on_audiocondition}, we compare our AudioComposer with several mainstream baseline models in terms of time, pitch and energy controllability on the AudioTPE test set. 
As shown in Table~\ref{res_on_cleanevent}, AudioComposer exhibits significant performance advantages in nearly all metrics. 
It is noteworthy that on energy control, AudioComposer exhibits a notable performance advantage in the $ACC$ metric, yet a inferior performance in the $MAE$ metric. 
This can be attributed to the input control information of our approach being limited to only three categories: high, normal, and low, rather than frame-level pitch and energy contours. 
As a result, the $ACC$ metric, which measures the accuracy of the three categories, demonstrates higher performance, while the frame-wise $MAE$ metric performance is less satisfactory. 
However, the superior category accuracy also indicates the AudioComposor can generate pitch and energy within more appropriate ranges than baselines.

\vspace{-1pt}
\subsection{Ablation Study}
\label{sec:ablation}
Table~\ref{res_on_ablation} provides further ablation studies on AudioCondition. 
Firstly, we remove flow matching and use the DDPM-based diffusion method, which results in a slight performance decline compared to AudioComposer-S. 
This suggests that the flow-based diffusion can enhance the model's control capability. 
In terms of inference efficiency, both DDPM-based diffusion and Tango need 200 steps, 
while our method can generate high-quality audio in just 25 steps. 
Our method also achieves better results with the same number of steps. 
Furthermore, when comparing our AudioComposer using DDPM-based diffusion transformer in Table~\ref{res_on_ablation} with Tango+LControl using U-Net diffusion from Table~\ref{res_on_audiocondition}, our diffusion transformer still achieves comparable performance, particularly in $F1_{event}$.

As shown in Table~\ref{res_on_ablation}, we remove the AudioCaps data and retrain the model AudioComposer-S, resulting in a performance decline. 
This validates that our method can utilize other coarse-grained data to enhance audio generation performance, as the models can be trained solely on NLDs in a holistic manner. 
Furthermore, AudioComposer can generate audio without style control, and we have evaluated it on the AudioCaps test set, as presented on our demo page due to page limitation.

% \footnote{https://lavendery.github.io/AudioComposer/}.

% Employing text solely for fine-grained control offers the benefit of also being able to utilize other coarse-grained data to enhance generation performance.

\section{conclusions}
\vspace{-1pt}
In this study, we present a fine-grained audio generation approach with natural language descriptions using flow-based diffusion transformers. The proposed method does not require additional conditions or complex network structures, as it relies solely on natural language descriptions to provide content specification and style control information with simplicity and efficiency.
We also propose a novel automatic data simulation pipeline that can construct fine-grained data and significantly alleviate the problem of data scarcity.   
Extensive experimental results prove that our approach enhances the speed, quality, and controllability of TTA generation and achieves state-of-the-art performances.

% \section*{Acknowledgment}

% The preferred spelling of the word ``acknowledgment'' in America is without 
% an ``e'' after the ``g''. Avoid the stilted expression ``one of us (R. B. 
% G.) thanks $\ldots$''. Instead, try ``R. B. G. thanks$\ldots$''. Put sponsor 
% acknowledgments in the unnumbered footnote on the first page.

\bibliographystyle{./IEEEtran}
\bibliography{./refs}

\begin{thebibliography}{10}
\providecommand{\url}[1]{#1}
\csname url@samestyle\endcsname
\providecommand{\newblock}{\relax}
\providecommand{\bibinfo}[2]{#2}
\providecommand{\BIBentrySTDinterwordspacing}{\spaceskip=0pt\relax}
\providecommand{\BIBentryALTinterwordstretchfactor}{4}
\providecommand{\BIBentryALTinterwordspacing}{\spaceskip=\fontdimen2\font plus
\BIBentryALTinterwordstretchfactor\fontdimen3\font minus \fontdimen4\font\relax}
\providecommand{\BIBforeignlanguage}[2]{{%
\expandafter\ifx\csname l@#1\endcsname\relax
\typeout{** WARNING: IEEEtran.bst: No hyphenation pattern has been}%
\typeout{** loaded for the language `#1'. Using the pattern for}%
\typeout{** the default language instead.}%
\else
\language=\csname l@#1\endcsname
\fi
#2}}
\providecommand{\BIBdecl}{\relax}
\BIBdecl

\bibitem{yang2023diffsound}
D.~Yang, J.~Yu, H.~Wang, W.~Wang, C.~Weng, Y.~Zou, and D.~Yu, ``Diffsound: Discrete diffusion model for text-to-sound generation,'' \emph{IEEE/ACM Transactions on Audio, Speech, and Language Processing}, 2023.

\bibitem{huang2023makeanaudio2}
J.~Huang, Y.~Ren, R.~Huang, D.~Yang, Z.~Ye, C.~Zhang, J.~Liu, X.~Yin, Z.~Ma, and Z.~Zhao, ``Make-an-audio 2: Temporal-enhanced text-to-audio generation,'' 2023.

\bibitem{guo2024audio}
Z.~Guo, J.~Mao, R.~Tao, L.~Yan, K.~Ouchi, H.~Liu, and X.~Wang, ``Audio generation with multiple conditional diffusion model,'' in \emph{Proceedings of the AAAI Conference on Artificial Intelligence}, vol.~38, no.~16, 2024, pp. 18\,153--18\,161.

\bibitem{kingma2018glow}
D.~P. Kingma and P.~Dhariwal, ``Glow: Generative flow with invertible 1x1 convolutions,'' \emph{Advances in neural information processing systems}, vol.~31, 2018.

\bibitem{goodfellow2020generative}
I.~Goodfellow, J.~Pouget-Abadie, M.~Mirza, B.~Xu, D.~Warde-Farley, S.~Ozair, A.~Courville, and Y.~Bengio, ``Generative adversarial networks,'' \emph{Communications of the ACM}, vol.~63, no.~11, pp. 139--144, 2020.

\bibitem{ho2020denoising}
J.~Ho, A.~Jain, and P.~Abbeel, ``Denoising diffusion probabilistic models,'' \emph{Advances in neural information processing systems}, vol.~33, pp. 6840--6851, 2020.

\bibitem{huang2023makeanaudio}
R.~Huang, J.~Huang, D.~Yang, Y.~Ren, L.~Liu, M.~Li, Z.~Ye, J.~Liu, X.~Yin, and Z.~Zhao, ``Make-an-audio: Text-to-audio generation with prompt-enhanced diffusion models,'' 2023.

\bibitem{liu2023audioldm}
H.~Liu, Z.~Chen, Y.~Yuan, X.~Mei, X.~Liu, D.~Mandic, W.~Wang, and M.~D. Plumbley, ``Audioldm: Text-to-audio generation with latent diffusion models,'' \emph{Proceedings of the International Conference on Machine Learning}, pp. 21\,450--21\,474, 2023.

\bibitem{ghosal2023texttoaudio}
D.~Ghosal, N.~Majumder, A.~Mehrish, and S.~Poria, ``Text-to-audio generation using instruction-tuned llm and latent diffusion model,'' 2023.

\bibitem{ghosal2023tango}
G.~Deepanway, M.~Navonil, M.~Ambuj, and P.~Soujanya, ``Text-to-audio generation using instruction-tuned llm and latent diffusion model,'' \emph{arXiv preprint arXiv:2304.13731}, 2023.

\bibitem{audioldm2-2024taslp}
H.~Liu, Y.~Yuan, X.~Liu, X.~Mei, Q.~Kong, Q.~Tian, Y.~Wang, W.~Wang, Y.~Wang, and M.~D. Plumbley, ``Audioldm 2: Learning holistic audio generation with self-supervised pretraining,'' \emph{IEEE/ACM Transactions on Audio, Speech, and Language Processing}, vol.~32, pp. 2871--2883, 2024.

\bibitem{majumder2024tango}
N.~Majumder, C.-Y. Hung, D.~Ghosal, W.-N. Hsu, R.~Mihalcea, and S.~Poria, ``Tango 2: Aligning diffusion-based text-to-audio generations through direct preference optimization,'' 2024.

\bibitem{kreuk2023audiogen}
F.~Kreuk, G.~Synnaeve, A.~Polyak, U.~Singer, A.~Défossez, J.~Copet, D.~Parikh, Y.~Taigman, and Y.~Adi, ``Audiogen: Textually guided audio generation,'' 2023.

\bibitem{yang2023uniaudio}
D.~Yang, J.~Tian, X.~Tan, R.~Huang, S.~Liu, X.~Chang, J.~Shi, S.~Zhao, J.~Bian, X.~Wu, Z.~Zhao, S.~Watanabe, and H.~Meng, ``Uniaudio: An audio foundation model toward universal audio generation,'' 2023.

\bibitem{yang2024uniaudio15}
D.~Yang, H.~Guo, Y.~Wang, R.~Huang, X.~Li, X.~Tan, X.~Wu, and H.~Meng, ``Uniaudio 1.5: Large language model-driven audio codec is a few-shot audio task learner,'' 2024.

\bibitem{xie2024picoaudio}
\BIBentryALTinterwordspacing
Z.~Xie, X.~Xu, Z.~Wu, and M.~Wu, ``Picoaudio: Enabling precise timestamp and frequency controllability of audio events in text-to-audio generation,'' 2024. [Online]. Available: \url{https://arxiv.org/abs/2407.02869}
\BIBentrySTDinterwordspacing

\bibitem{lipman2022flow}
Y.~Lipman, R.~T. Chen, H.~Ben-Hamu, M.~Nickel, and M.~Le, ``Flow matching for generative modeling,'' \emph{arXiv preprint arXiv:2210.02747}, 2022.

\bibitem{peebles2023scalable}
W.~Peebles and S.~Xie, ``Scalable diffusion models with transformers,'' in \emph{Proceedings of the IEEE/CVF International Conference on Computer Vision}, 2023, pp. 4195--4205.

\bibitem{esser2024scaling}
P.~Esser, S.~Kulal, A.~Blattmann, R.~Entezari, J.~M{\"u}ller, H.~Saini, Y.~Levi, D.~Lorenz, A.~Sauer, F.~Boesel \emph{et~al.}, ``Scaling rectified flow transformers for high-resolution image synthesis,'' in \emph{Forty-first International Conference on Machine Learning}, 2024.

\bibitem{gao2024lumi}
\BIBentryALTinterwordspacing
P.~Gao, L.~Zhuo, D.~Liu, R.~Du, X.~Luo, L.~Qiu, Y.~Zhang, C.~Lin, R.~Huang, S.~Geng, R.~Zhang, J.~Xi, W.~Shao, Z.~Jiang, T.~Yang, W.~Ye, H.~Tong, J.~He, Y.~Qiao, and H.~Li, ``Lumina-t2x: Transforming text into any modality, resolution, and duration via flow-based large diffusion transformers,'' 2024. [Online]. Available: \url{https://arxiv.org/abs/2405.05945}
\BIBentrySTDinterwordspacing

\bibitem{kingma2013auto}
D.~P. Kingma and M.~Welling, ``Auto-encoding variational bayes,'' \emph{arXiv preprint arXiv:1312.6114}, 2013.

\bibitem{raffel2020exploring}
C.~Raffel, N.~Shazeer, A.~Roberts, K.~Lee, S.~Narang, M.~Matena, Y.~Zhou, W.~Li, and P.~J. Liu, ``Exploring the limits of transfer learning with a unified text-to-text transformer,'' \emph{Journal of machine learning research}, vol.~21, no. 140, pp. 1--67, 2020.

\bibitem{lee2022bigvgan}
S.-g. Lee, W.~Ping, B.~Ginsburg, B.~Catanzaro, and S.~Yoon, ``Bigvgan: A universal neural vocoder with large-scale training,'' \emph{arXiv preprint arXiv:2206.04658}, 2022.

\bibitem{simplespeech}
D.~Yang, D.~Wang, H.~Guo, X.~Chen, X.~Wu, and H.~Meng, ``Simplespeech: Towards simple and efficient text-to-speech with scalar latent transformer diffusion models,'' \emph{Proc. INTERSPEECH}, 2024.

\bibitem{simplespeech2}
D.~Yang, R.~Huang, Y.~Wang, H.~Guo, D.~Chong, S.~Liu, X.~Wu, and H.~Meng, ``Simplespeech 2: Towards simple and efficient text-to-speech with flow-based scalar latent transformer diffusion models,'' 2024.

\bibitem{podell2023sdxl}
D.~Podell, Z.~English, K.~Lacey, A.~Blattmann, T.~Dockhorn, J.~M{\"u}ller, J.~Penna, and R.~Rombach, ``Sdxl: Improving latent diffusion models for high-resolution image synthesis,'' \emph{arXiv preprint arXiv:2307.01952}, 2023.

\bibitem{su2024roformer}
J.~Su, M.~Ahmed, Y.~Lu, S.~Pan, W.~Bo, and Y.~Liu, ``Roformer: Enhanced transformer with rotary position embedding,'' \emph{Neurocomputing}, vol. 568, p. 127063, 2024.

\bibitem{zhang2023llama}
R.~Zhang, J.~Han, C.~Liu, P.~Gao, A.~Zhou, X.~Hu, S.~Yan, P.~Lu, H.~Li, and Y.~Qiao, ``Llama-adapter: Efficient fine-tuning of language models with zero-init attention,'' \emph{arXiv preprint arXiv:2303.16199}, 2023.

\bibitem{gao2023llama}
P.~Gao, J.~Han, R.~Zhang, Z.~Lin, S.~Geng, A.~Zhou, W.~Zhang, P.~Lu, C.~He, X.~Yue \emph{et~al.}, ``Llama-adapter v2: Parameter-efficient visual instruction model,'' \emph{arXiv preprint arXiv:2304.15010}, 2023.

\bibitem{zhang2023adding}
L.~Zhang, A.~Rao, and M.~Agrawala, ``Adding conditional control to text-to-image diffusion models,'' in \emph{Proceedings of the IEEE/CVF International Conference on Computer Vision}, 2023, pp. 3836--3847.

\bibitem{fonseca2021fsd50k}
E.~Fonseca, X.~Favory, J.~Pons, F.~Font, and X.~Serra, ``Fsd50k: an open dataset of human-labeled sound events,'' \emph{IEEE/ACM Transactions on Audio, Speech, and Language Processing}, vol.~30, pp. 829--852, 2021.

\bibitem{piczak2015dataset}
K.~J. Piczak, ``{ESC}: {Dataset} for {Environmental Sound Classification},'' in \emph{Proceedings of the 23rd {Annual ACM Conference} on {Multimedia}}.\hskip 1em plus 0.5em minus 0.4em\relax {ACM Press}, 2015, pp. 1015--1018.

\bibitem{salamon2014dataset}
J.~Salamon, C.~Jacoby, and J.~P. Bello, ``A dataset and taxonomy for urban sound research,'' in \emph{Proceedings of the 22nd ACM international conference on Multimedia}, 2014, pp. 1041--1044.

\bibitem{45857}
J.~F. Gemmeke, D.~P.~W. Ellis, D.~Freedman, A.~Jansen, W.~Lawrence, R.~C. Moore, M.~Plakal, and M.~Ritter, ``Audio set: An ontology and human-labeled dataset for audio events,'' in \emph{Proc. IEEE ICASSP 2017}, New Orleans, LA, 2017.

\bibitem{9414579}
S.~Hershey, D.~P.~W. Ellis, E.~Fonseca, A.~Jansen, C.~Liu, R.~Channing~Moore, and M.~Plakal, ``The benefit of temporally-strong labels in audio event classification,'' in \emph{ICASSP 2021 - 2021 IEEE International Conference on Acoustics, Speech and Signal Processing (ICASSP)}, 2021, pp. 366--370.

\bibitem{audiocaps}
C.~D. Kim, B.~Kim, H.~Lee, and G.~Kim, ``Audiocaps: Generating captions for audios in the wild,'' in \emph{NAACL-HLT}, 2019.

\bibitem{kingma2014adam}
D.~P. Kingma, ``Adam: A method for stochastic optimization,'' \emph{arXiv preprint arXiv:1412.6980}, 2014.

\bibitem{ebbers2022pre}
J.~Ebbers and R.~Haeb-Umbach, ``Pre-training and self-training for sound event detection in domestic environments,'' 2022.

\bibitem{mesaros2016metrics}
A.~Mesaros, T.~Heittola, and T.~Virtanen, ``Metrics for polyphonic sound event detection,'' \emph{Applied Sciences}, vol.~6, no.~6, p. 162, 2016.

\bibitem{ren2020fastspeech}
Y.~Ren, C.~Hu, X.~Tan, T.~Qin, S.~Zhao, Z.~Zhao, and T.-Y. Liu, ``Fastspeech 2: Fast and high-quality end-to-end text to speech,'' \emph{arXiv preprint arXiv:2006.04558}, 2020.

\end{thebibliography}

\end{document}